# Interference mitigation techniques for a dense heterogeneous area network in machine-to-machine communications

Dong Chen[1*], Jamil Khan[1], Muhammad Awais Javed[2], Jason Brown[3]

1. The University of Newcastle, Electrical Engineering and Computing, University Drive, 2308, NSW, Australia

2. COMSATS University Islamabad, Pakistan

3. The University of Southern Queensland (USQ) 37 Sinnathamby Blvd, Springfield Central QLD, 4300, Australia

**Abstract**:

With the advent of Machine-to-Machine (M2M) communications, various networking consumer industrial and autonomous systems exchange messages in the real world in order to achieve their objectives. Parts of these systems are comprised of short-range wireless networks in the form of clusters that collectively cover a large geographical area. In these clusters, the nodes that represent the cluster heads need to deal with two types of communications: one is within the cluster and the other is from the cluster to the sink node. As the number of clusters increases, it takes multiple hops for the cluster head to forward data to the sink node, thus resulting in a low packet delivery rate and throughput. To solve this problem, we propose a heterogeneous area network in which the cluster head is equipped with two types of radios: the IEEE 802.15.4 and IEEE 802.11 radios. The former is for the devices within the cluster to communicate, whereas the latter is for the cluster heads to communicate to the sink node. Although the IEEE 802.11 links increase the link capacity, the IEEE 802.11 radio and the IEEE 802.15.4 radio might share the 2.4 GHz unlicensed band, thus giving rise to the inter-network collisions or interference. To tackle this problem and to maintain decent Quality-of-Service (QoS) for the network, we subsequently present two interference mitigation techniques, in which a Blank Burst (BB) period is proposed so that the IEEE 802.15.4 radios can be suspended while the IEEE 802.11 radios are active. Simulation results show the proposed two methods can effectively mitigate the inter-network collisions and are superior to the existing technique, which uses an adaptive aggregation technique to mitigate the inter-network collisions.

## 1 Introduction

Wireless Sensor Networks (WSNs) have gained much attention from academia in fields such as environmental monitoring and ubiquitous computing [1]. Most WSNs are based on the short-range IEEE 802.15.4 standard, which can be characterized as a low data rate, low energy consumption and low cost specification. The IEEE 802.15.4 standard can also provide seamless connectivity to sensors within a short range. However, WSNs normally need to cover a large-scale area in which sensor nodes and the sink node can be far from each other. A widely accepted solution is clustering; that is, cluster heads are responsible for wirelessly connecting to the sink node while sensors communicate with their local cluster head, as shown in Fig. 1. Note that some cluster heads are located at the edge of the area and might need a multi-hop connection to the data sink. Several studies have proved that a multi-hop WSN experiences a low packet delivery rate and a high end-to-end delay [2]. This is because the accumulated traffic is beyond the capacity of the relay cluster heads, thus resulting in buffer overflow and the significant performance degradation of the WSN.

To tackle this issue, we connect the cluster head to an IEEE 802.11 (i.e., Wi-Fi) node to form a heterogeneous area network. The IEEE 802.11 standard can be characterised as a high data rate and high throughput specification, which is good compensation for the shortcomings of IEEE 802.15.4 networks. In this work, we adopt IPv6 over Low-Power Wireless Personal area network (6LoWPAN) networks as the networks based on the IEEE 802.15.4 standard. In the proposed heterogeneous area network, the IEEE 802.11 network serves as the backbone network connecting the cluster heads and the sink node. More specifically, IEEE 802.11 networks have a higher transmission rate up to 400 Mbps (e.g., IEEE 802.11n) and a much longer transmission range up to 300 metres compared to 6LoWPAN networks. As a result, these two advantages can help the 6LoWPAN networks handle a large amount of traffic over a large-scale area. It is noted that both the two networks share the unlicensed 2.4 GHz band, so the relatively high transmission power of the IEEE 802.11 network can adversely affect the transmissions of the 6LoWPAN network and cause inter-network collisions that lower the packet delivery ratio and increase the end-to-end delay.

---

*Corresponding author: Dong.Chen@uon.edu.au

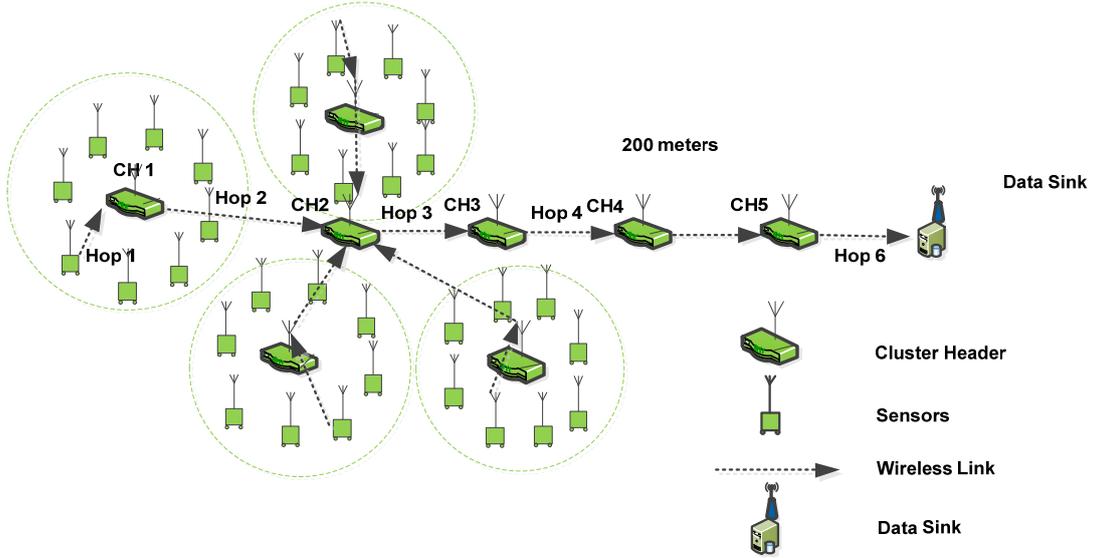

Fig. 1 A multi-hop wireless sensor network

In this paper, we propose two techniques to mitigate the inter-network collisions: an aggregation factor-based algorithm and a lifetime-based algorithm. The latter is an enhanced version of the former and can offer better QoS for the network performance. The main contributions of this paper are as follows:

- In the first proposed algorithm, the cluster head transmits aggregated 6LoWPAN packets in the form of WLAN packets while suspending the 6LoWPAN transmissions for a BB period to avoid the inter-network collisions. To trigger the algorithm, an aggregation factor is adopted to count the number of 6LoWPAN packets to be aggregated. Every time the number of 6LoWPAN packets equals the aggregation factor, the algorithm is triggered.

- The second proposed algorithm is a lifetime-based algorithm in which the cluster head also stops the 6LoWPAN transmissions during a BB period, but this time with a different triggering mechanism. The algorithm keeps a record of the shortest lifetime (i.e., the time for which a packet is considered useful to the receiver) in the network. Once this lifetime is equivalent to a predefined safe lifetime margin that guarantees that all the transmitted packets do not expire based on their lifetimes, the algorithm is triggered to maintain the QoS of the corresponding M2M applications. As part of the process, the aggregation factor is also used to control the number of the IEEE 802.11 packets.

- To show the superiority of the proposed two algorithms, they are compared with existing methods from the literature. Due to the introduced BB period, the proposed two algorithms can more effectively mitigate the inter-network collisions than the existing work.

The remainder of this paper is organized as follows: the related work is presented in Section 2. The development of the proposed heterogeneous area network is described in Section 3. Sections 4 and 5 propose two interference mitigation techniques, which are the aggregation factor-based and lifetime-based algorithms. The performances of the proposed algorithms are discussed in Section 6. Section 7 concludes the paper.

## 2 Related work

Research studies on interference between IEEE 802.15.4 and IEEE 802.11 networks have attracted great attention in recent years. Many researchers analysed the impact of the interference between the two separate networks. Mahmoud Talebi et.al [3] analysed the performances of IEEE 802.15.4 networks in the presence of WLAN interference. It was found that IEEE 802.15.4 networks are more susceptible to log-normal WLAN traffic, instead of regular periodic WLAN traffic. The analysis plays an important role in understanding the interference in large IoT networks. N. de A. Moreira. et.al [4] concluded that ZigBee nodes are efficient in listening a WLAN receiver, and differences in the physical layer and the MAC layer in terms of symbol duration, carrier frequencies and modulation schemes, make it difficult for the ZigBee nodes to avoid interference. R. Casagrande et.al [5] analysed the interference between IEEE 802.11 n and IEEE 802.15.4 networks. It was concluded that if the 802.15.4 coordinator is closer to the WLAN station, the packet loss rate and the end-to-end delay become higher. In addition, if the WLAN channel occupancy rate increases, the packet loss rate and the end-to-end delay significantly increase, as well as the decrease in the receive signal strength indicator (RSSI). The impact of WLAN nodes on ZigBee nodes cannot be overlooked. Sharad et.al [6] analysed the IEEE 802.15.4 network performance in the presence of WLAN nodes. It was concluded that the safe distance between the two networks is 8 meters and the safe frequency offset is 8 MHz when the distance is just 2 meters. Jin Seok Han et.al [7] investigated how IEEE 802.15.4 networks are affected by WLAN networks when sharing the 2.4 GHz band. The probabilities of transmission failure and data throughput under the interference were modelled using a Markov chain model. An analytical model was made to find out the optimum packet size to maximise the data throughput as per the interference channel features.

In addition, some frameworks have been proposed in the smart home environment. Vikram.K and Sarat Kumar Sahoo [8] proposed a collaborative framework to mitigate the interference between the ZigBee and the WLAN nodes in a smart home environment. The ZigBee nodes assess the channel conditions, delay and throughput. If these metrics cannot maintain the basic ZigBee transmissions, an alert-control signal is transmitted to the ZigBee coordinator that in turn relays the signal to a home gateway connecting both the WLAN and ZigBee networks. Then the signal is received by the WLAN access points that later reduce the traffic to avoid the interference. Kunho Hong et.al [9] proposed two algorithms that help the ZigBee nodes to keep the maximum delay allowed by applications while maintaining high throughput of the WLAN nodes, in the presence of the WLAN interference. These two algorithms ask to the WLAN nodes to stop transmissions for a short period of time to avoid the interference with the ZigBee nodes. Shu Nishikori et.al [10] proposed a cooperative channel control method of ZigBee and WLAN to reduce the level of the interference. The proposed method requires the WLAN to stop for a short time period. Meanwhile, the ZigBee nodes switch to a more suitable channel to avoid the interference, and the length of pausing for WLAN nodes is limited to maintain the WLAN throughput. Hao Ran Chi et.al [11] proposed a multi-objective optimization interference mitigation ZigBee model for Advanced Metering Infrastructure. When the ZigBee nodes detect the interference from the WLAN nodes, a channel swapping system implemented in in the infrastructure helps the ZigBee nodes to change the channel to avoid the interference from the WLAN nodes. Fumihiro et al [12] proposed a hybrid station. One part of this station is the WPAN node, whereas the other part is the WLAN node. To avoid the interference between the two networks, the hybrid station schedules the transmissions of the two networks to use the Network Allocation Vector (NAV) to transmit WPAN packets.

The above studies are all based on two separate networks, which means IEEE 802.15.4 and IEEE 802.11 networks are two independent networks. Additionally, there are also some hybrid networks proposed in the literature. The notion of large-scale dual-radio wireless sensor networks was proposed by Anis Koubaa et al [13]. The energy-constrained large-scale sensor network is connected with WLAN networks, which have a high data rate and a long transmission range. This two-tiered network can improve reliability, scalability and the real-time QoS performance. Specifically, the high data rate and long transmission range of WLAN networks allow real-time transmissions of the sensor traffic with the required QoS. WLANs are less susceptible to the inter-network collisions than the sensor nodes, so WLANs can be used as backbone networks in this work. Wireless sensor networks often need to increase the number of nodes to cover a wide area, so WLANs can be used to extend the transmission range by forming WLAN mesh networks. However, the work in [13] provided a conceptual design without detailed results. Similar ideas were also proposed in [14, 15]. A sensor network for IP-WSN gateway was developed in [16]. In this work, an IP-based sensor network connects WLAN networks to make the network globally accessible to the Internet in a fast and cost-effective manner. Although this work implemented a hardware platform using TI CC2520 devices, it did not provide any test plans or simulation results.

A dual-radio network was used in the building blocks in [17]. This work implemented a WiFi-ZigBee hybrid building sensor network to tackle the problem of efficiently deploying Advanced Metering Infrastructure (AMI) with heavy network loads and in the difficult radio propagation environment. To investigate the hybrid network performance, a case study was conducted. The simulation results of this work showed that the round trip time for the demand response applications was 0.6s and that the one-way time transmission for smart metering is around 9s. The ZigBee nodes connect to a dual-radio node using only one hop. The work did not report if the ZigBee network could be affected by the WLAN transmissions. Jun Wang et al [18] proposed a heterogeneous ZigBee/Wi-Fi network and compared the network performance with that of ZigBee networks in terms of throughput, packet loss rate, the packet loss ratio and average end-to-end delay using the OPNET modeller. It was found that the hybrid technique outperformed the ZigBee network and had a good performance at lower traffic loads. However, the hybrid technique adopted a star topology and did not take the inter-network collisions into account. Another study used a ZigBee/WLAN dual-radio node to form a multi-tier multi-hop heterogeneous sensor network monitoring transportation networks such as trains and truck platoons [19]. Specifically, the sensor nodes were organised into clusters and wirelessly connected with the other nodes in the same cluster using the ZigBee radio. The communications between the clusters was enabled by a ZigBee/Wi-Fi gateway with a Wi-Fi radio. The work used the OPNET simulation model and the simulation results were compared with the theoretical analysis, but the Wi-Fi and ZigBee radios used two different channels to avoid the inter-network collisions.

The only work that considers the inter-network collisions was proposed in [20]. In this work, a dual-radio network based on the IEEE 802.11 and IEEE 802.15.4 standards was proposed to extend the effective transmission range of the IEEE 802.15.4 multi-hop network. It was expected that this extension might create the inter-network collisions in the IEEE 802.15.4 networks. As a result, an adaptive aggregation technique was proposed to reduce the number of IEEE 802.11 packets, thus mitigating the inter-network collisions. To fully take advantage of the IEEE 802.11 payload, 25 6LoWPAN payloads were aggregated into one IEEE 802.11 packet. Reducing the number of 6LoWPAN packets proved to be effective in mitigating the inter-network collisions. In this paper, this work is used to compare with our proposed algorithms in Section 6. The difference between this work and our proposed methods is that our methods have a short period to protect the IEEE 802.15.4 packets, whereas the existing work does not. We take advantage of the WLAN high transmission rate that is hundred times faster than that of the IEEE 802.15.4 networks, so the short period is small enough so that it would not affect the 6LoWPAN throughput.

## 3 Heterogeneous wireless area network design

### 3.1 Heterogeneous wireless area network architecture

To cover a large-scale area, wireless area networks must be scalable in increasing the number of devices, so clustering is a feasible and widely accepted method for network scalability. In the large-scale wireless network, changing the operational channel from a busy one to an idle one may not be a feasible solution to mitigate the inter-network collisions. The reasons are twofold. The first is that informing all of the devices in the network to switch the channel could incur huge communication overheads for those devices; the second is that it may be difficult to locate a free channel due to the density of interfering devices. For example, if a sports stadium deploys a ZigBee-based lighting control system, it might be interfered by the audience using cell phones or laptops based on the IEEE 802.11 standard.

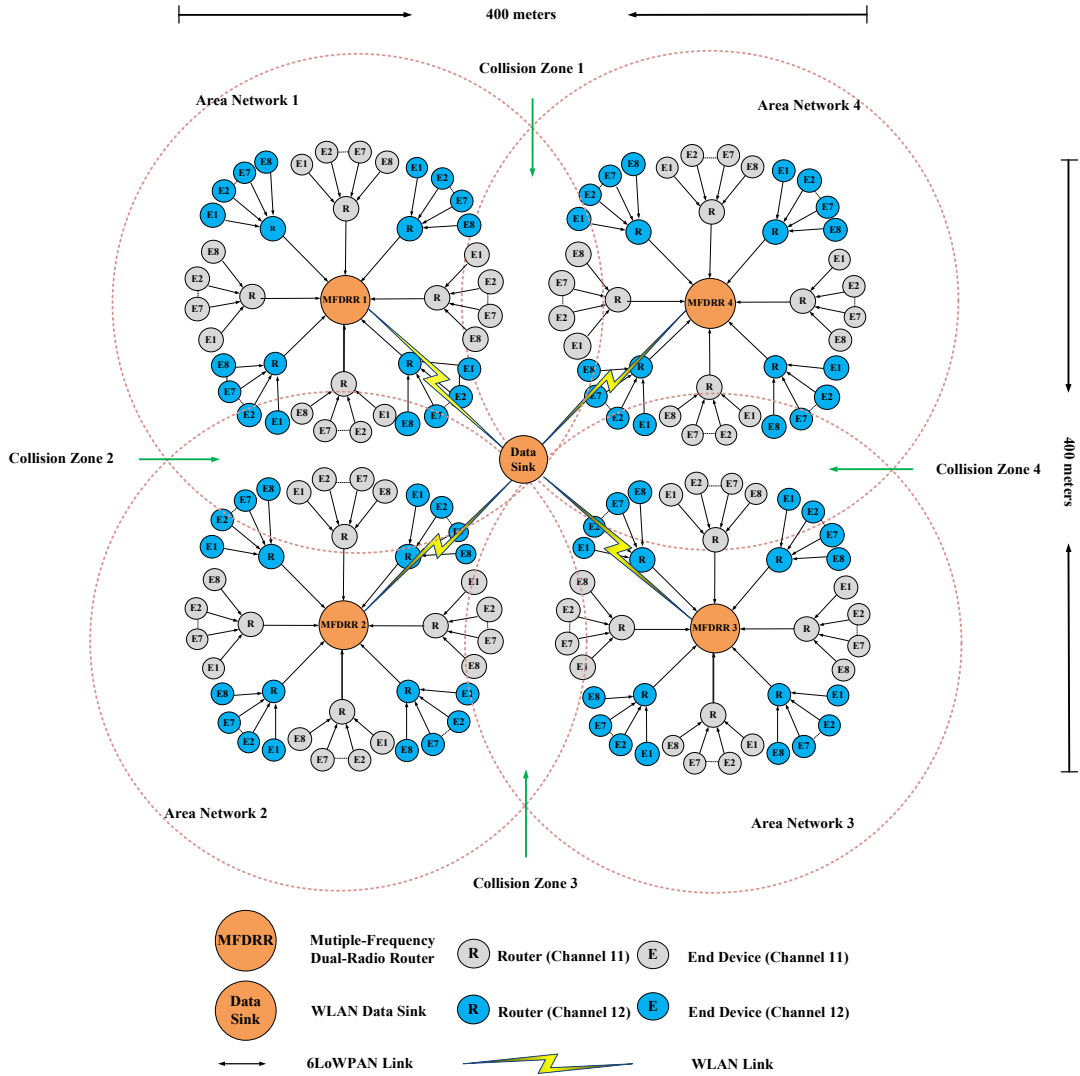

Fig. 2 The large-scale heterogeneous wireless sensor architecture

Figure 2 depicts the heterogeneous wireless area network architecture considered in this paper. This figure has extended the network layout presented in [20], in which a two-hop network composed of nodes with CC2420 microprocessors running Tiny OS [21] and Linux was adopted. The routers proposed in [20], namely clusters, used the Marvell PXA320 microprocessor. The CC2420 chipset was used for the IEEE 802.15.4 radio, while the Marvell 88W8686 chipset was used for IEEE 802.11 b radio in the literature. In contrast, in our proposed network, the 6LoWPAN and IEEE 802.11g standards are considered. It can be seen that all the end devices are governed by the routers. The routers are responsible for communications within the cluster and for communications from the router to the PAN coordinator, which is a Multi-Frequency Dual-Radio Router (MFDRR). The MFDRR then forwards the collected data to the data sink. The MFDRR has two protocol stacks: the 6LoWAN stack and the WLAN stack. The 6LoWPAN stack supports two frequencies represented in blue and grey in the network. Fig. 2 also presents four small area networks, each of which contains 64 6LoWPAN nodes, eight routers, so there are totally 293 nodes covering a 400×400 $m^2$ area, which is similar to the size of a schoolyard.

This large-scale area network can be extended to several $km^2$ by using the spatial reuse technique, so thousands of 6LoWPAN devices running various M2M applications can be incorporated in a large-scale geographical area network. The uplink data flow of the proposed network goes from the

6LoWPAN end devices to the data sink via the 6LoWPAN routers and the MFDRRs. Note that the four MFDRRs create four inter-network collision zones. This means one MFDRR not only affect its own area network, but also affects its two neighbouring area networks. As the four MFDRRs start to transmit, the level of the inter-network collision increases, so the performance of the network would be affected. The negative effects tend to be exacerbated in such a large-scale heterogeneous area network. The performances of the algorithms are evaluated in a large-scale area network with four MFDRRs using the simulation model.

The IEEE 802.15.4 and IEEE 802.11 standards support multiple transmission frequency bands including the 2.4 GHz ISM band. This paper focuses on the 2.4 GHz transmission bands where the transmission channels of the 6LoWPAN and MFDRR overlap. The solutions developed in this work can be extended to other frequency bands, particularly the 900MHz frequency band. Fig. 3 shows the sub-channel arrangement of the IEEE 802.11 and IEEE 802.15.4 standards in the 2.4 GHz frequency spectrum. The 802.11b/g standards use a 20 MHz transmission bandwidth, while 6LoWPAN uses a narrow 2MHz transmission band. The figure shows that the 1, 6 and 13 WLAN channels significantly interfere with 802.15.4 channels, and each WLAN channel interferes with four 6LoWPAN channels. The WLAN packets are also transmitted with a much higher power level than the 6LoWPAN because of the higher transmission distance involved, resulting in a longer interference distance to IEEE 802.15.4 devices. In the dense networking scenario, the interference could be worse due to the co-location of many MFDRRs and 6LoWPAN devices.

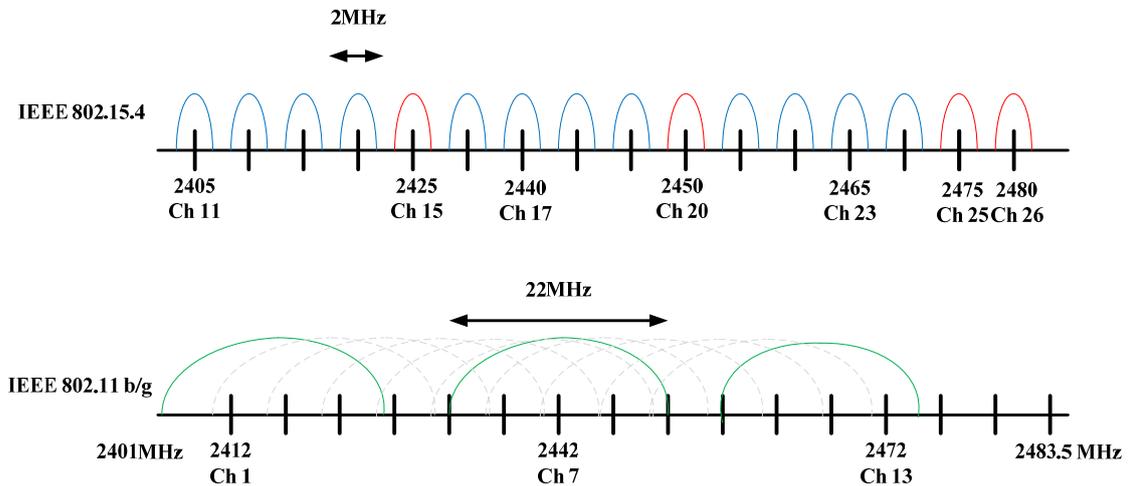

Fig. 3 Transmission channels of the IEEE 802.15.4 and IEEE 802.11 standards in the 2.4 GHz band

### 3.2 6LoWPAN Node Model

The Optimised Networking Engineering Tool (OPNET) is a hierarchical modelling tool and consists of three domains: the network domain, the node domain and the process domain, as shown in Fig. 4. The network domain regulates the topology in which the nodes and communication links are properly organized. The node domain represents the different nodes that are the basic communication entities in the network domain and allows packets to flow through different modules. The process domain is the lowest level of simulation, in which all the functionalities and connections are implemented in Proto C language. A process model is comprised of a finite state machine, in which each state represents a logical operation carried out on data and each condition triggers the execution of code in each state. Unfortunately, no 6LoWPAN model is available in OPNET modeller, so we developed the model from scratch. Thanks to the implementation of the IEEE 802.15.4 MAC layer of open-zb [22], the 6LoWPAN model can be developed using the open-zb model and other models from the OPNET model library. Open-zb, based on the IEEE 802.15.4 standard and ZigBee specifications, is open-source, so developing the 6LoWPAN node on top of this model is an ideal choice. The detailed node model development processes are illustrated below.

In addition, before implementing a 6LoWPAN node model, the internal model structures, such as the modules and packet streams, need to be carefully considered. As illustrated in Fig. 5, the

6LoWPAN node model has been implemented as per the 6LoWPAN protocol stack and combines two different node models. The first built-in IP layer model is truncated from an Ethernet node model named Ethernet_ ip_station_adv from the standard OPNET library; the MAC layer model is obtained from the open-zb node model. The adaptation and the upper layer models were developed from scratch. In particular, since the IEEE 802.15.4 standard defines three types of devices i.e., PAN coordinator, Full Function Device (FFD) and Reduced Function Device (RFD), the corresponding devices (i.e., PAN Coordinator, Router and End Device), complying with the 6LoWPAN protocol, were also developed to set up the network scenarios in this work. To this end, numerous changes were made to meet the requirements of the 6LoWPAN model. Since the Ethernet model has an Address Resolution Protocol (ARP) layer located in the middle of a MAC and IP layer, how the ARP layer interfaces with the IP layer should be thoroughly investigated. Similarly, the open-zb node model has a network layer directly above its MAC layer, so it is necessary to understand how the network layer interfaces with the IEEE 802.15.4 MAC layer.

The two patterns reveal how an adaptation layer can be implemented as per the connection patterns in open-zb and the Ethernet node model. Moreover, the 6LoWPAN application layer node model is designed to interface with the IP layer and to generate the different types of M2M traffic. The 6LoWPAN application layer implementation is referenced from the open-zb application layer model in terms of setting the Interface Control Information (ICI) and traffic profiles. The open-zb application layer explains more details on how the open-zb node application generates traffic to the lower layer, so this pattern can be applied to the 6LoWPAN node model sending packet flows down to the IP layer.

The IP layer is the most complicated model in the OPNET library and consists of many functions such as TCP/IP protocol suite, dual-stack IPv4/IPv6, routing, ARP, fragmentation and Network Address Translation (NAT), as well as the routing policy and firewall filters. All these functions are integrated into these two modules sharing the same code operating in different modes. The OPNET document (OPNETWORKS) suggests that no modules should be added between the IP module and ARP module; otherwise, it would be difficult to debug the code [23]. However, to interface the IP module with the open-zb MAC and PHY layer, the IP module needs to be separated from the ARP model as shown in Fig. 5. This process proved to be intricate, especially connecting the two modules in OPNET via debugging. Accordingly, four packet streams for the adaptation layer are defined in the Header Block.

STRM_FROM_IP_TO_LoWPAN

STRM_FROM_LoWPAN_To_IP

STRM_FROM_MAC_To_LoWPAN

STRM_FROM_LoWPAN_To_MAC

These streams help transfer data packets between the IP layer and MAC layer. Another two packet streams connect the application layer and IP layer, and provide communication between these two layers.

STRM_FROM_IP_ENCAP_TO_ APP

STRM_FROM_APP_TO_IP_ENCAP

Moreover, the attributes were created for the simulation configuration such that they can be promoted to the network level to enable multiple runs. For example, the open-ZB node model supports the acknowledged and non-acknowledged traffic. To keep these functionalities, a corresponding change was made in the adaptation layer to control the ACK packets of the traffic. Many other attributes, such as IPv6 prefixes, were also integrated into the node model so that the stateless address auto-configuration can be used to generate IPv6 addresses. A number of other modifications were also made to facilitate OPNET simulation described below.

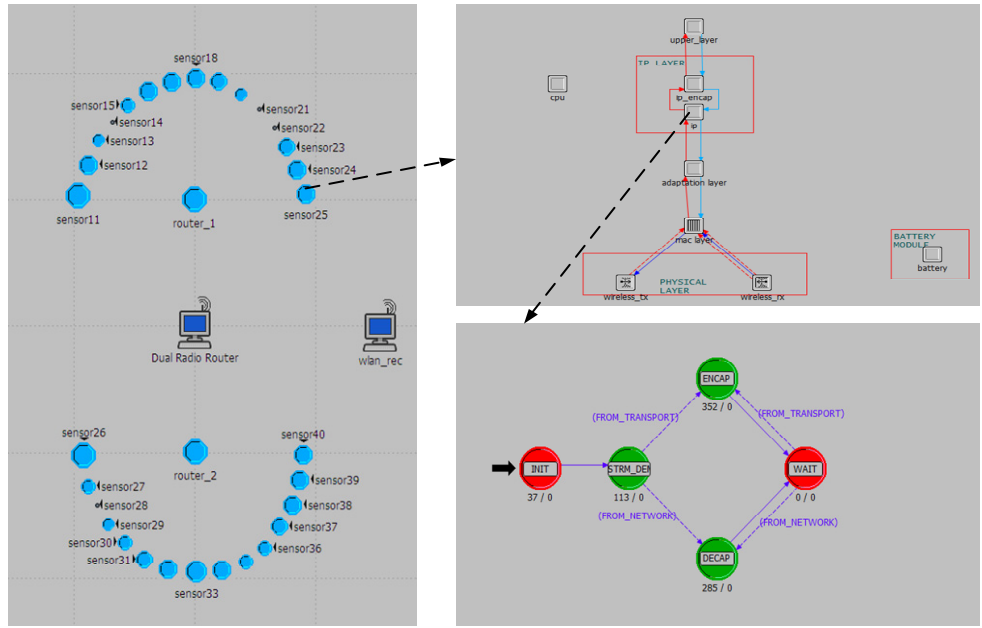

Fig. 4 The three OPNET domains (left: network, top right: node, bottom right: process)

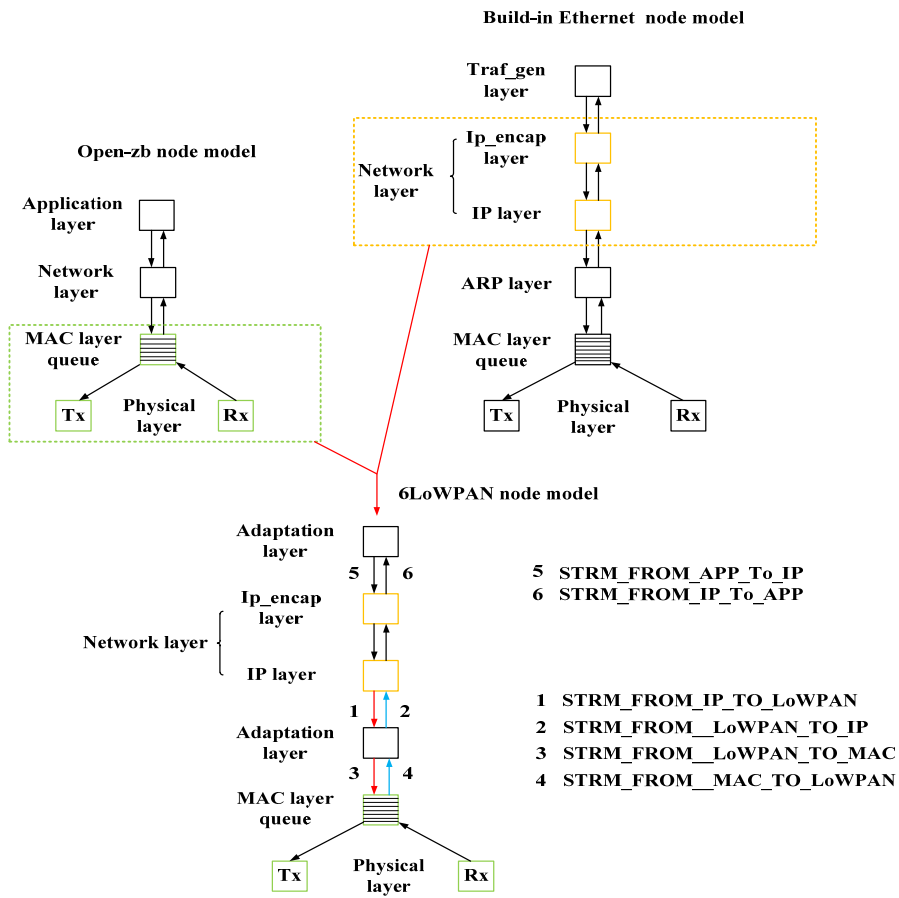

Fig. 5 6LoWPAN OPNET node model and stream flows

### 3.3 MFDRR Node Mode

The algorithm discussed in this work is located in the application layer of the dual-protocol stack, as shown in Fig. 6. It shows the two protocol stacks of the MFDRR: the 6LoWPAN stack and WLAN stack. In addition, packet processing within the MFDRR begins in the following way: upon receiving a sensor packet from the 6LoWPAN MAC layer, the 6LoWPAN stack strips off headers and forwards payloads to the WLAN application layer. All the payloads are stored in the aggregation buffer that aggregates them into a WLAN payload for further transmissions. The 6LoWPAN stack uses two MAC layers and two physical interfaces that can support two frequencies; the WLAN stack uses a fixed channel that covers two channels used by the 6LoWPAN devices, as shown in Fig. 7.

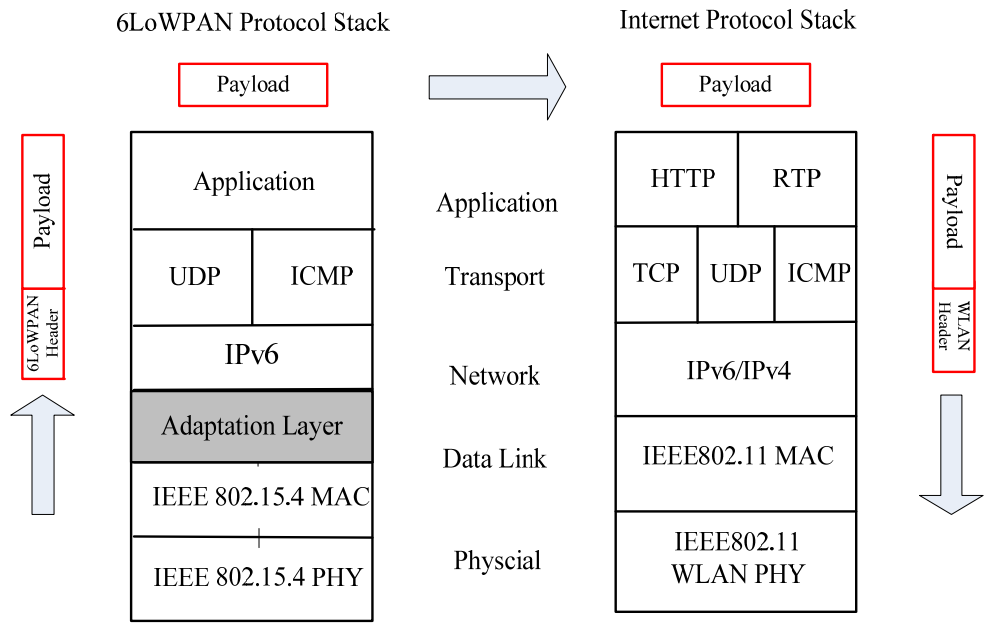

Fig. 6 MFDRR protocol stack

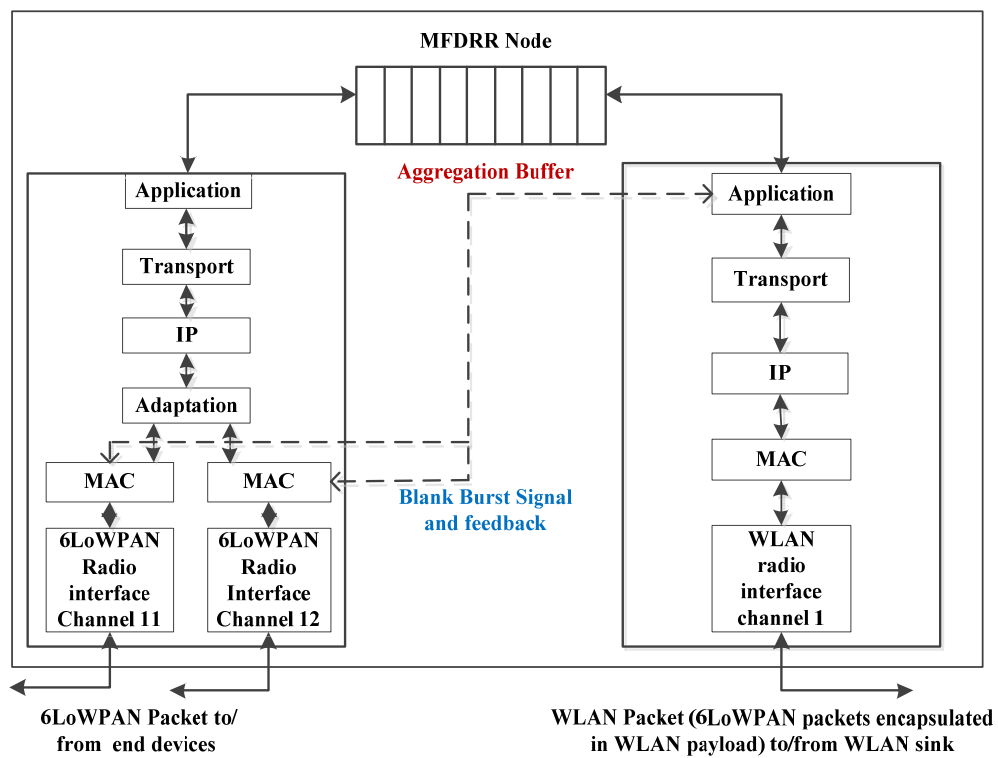

Fig. 7 MFDRR protocol with two 6LoWPAN frequencies

# 4 Aggregation-factor-based BB algorithm design

The WLAN transmissions adversely affect the 6LoWPAN transmissions due to the inter-network collisions. The reason is that when the MFDRR will transmit the WLAN packets to the data sink. the WLAN packets will interfere with the 6LoWPAN nodes especially the routers due to the broadcast nature of wireless communication. More precisely, the WLAN signals will mix with the 6LoWPAN signals in the physical layer and increase the noise level, thus lowering the Signal-to-Noise Ratio (SNR) for the 6LoWPAN packets that will be discarded by the receiver. We can express the throughput affected by the inter-network collision as follows.

$$Thr_1 * (1 - \alpha) = Thr_2 \ (0 < \alpha < 1), \tag{1}$$

where $Thr_1$ denotes the throughput of the MFDRR without the adverse impact of the inter-network collisions, $\alpha$ denotes a coefficient representing to what extent $Thr_1$ can be interfered with by the inte-network collisions, $Thr_2$ denotes the affected throughput. In this paper, unlike the two separate networks described in the related work, the proposed heterogeneous area network is one network in which the WLAN interface of the MFDRR interferes with the 6LoWPAN nodes all the time. This is because the 6LoWPAN nodes are the data source of the MFDRR, so once the 6LoWPAN is transmitting, the WLAN interface is transmitting as well. As it is difficult to shift channels in a crowded surrounding as discussed in section 3.1, we adopt the time domain and attempt to transmit WLAN packets in an exclusive period without other 6LoWPAN transmissions. Under this condition, $\alpha$ approximates to zero in (1), then $Thr_2$ approximates to $Thr_1$, but with the cost of higher delays that equals to the exclusive period. This period is named as Blank Burst (BB) to suspend the 6LoWPAN transmissions from the MAC layer as the WLAN packet transmissions start. Before the 6LoWPAN end devices stop transmitting packets, a BB signal is piggybacked in a beacon, which is propagated to end devices indicating that the devices should suspend their transmissions for a short period of Blank Burst time. In doing so, the WLAN transceivers of the MFDRR use this silent period to transmit WLAN traffic without interfering with 6LoWPAN transmissions. Therefore, the performance gain G obtained by using the proposed techniques can be expressed as below.

$$G = Thr_1 - Thr_2 = Thr_1 - Thr_1 * (1 - \alpha) = \alpha Thr_1 > 0 \tag{2}$$

Because $0 < \alpha < 1$ and $Thr_1 > 0$, so the value of G is also above zero, which proves the advantage of the proposed technique.

As shown in Fig. 8, the BB signal is illustrated to solve the inter-network collisions. When the 6LoWPAN packets arrive at the buffer in the MFDRR as shown in Fig. 7, they are aggregated into WLAN packets. Once the number of packets in the buffer exceeds a pre-defined aggregation factor threshold, a BB signal request is triggered at time T1 and sent from the WLAN application layer to the 6LoWPAN MAC layer, prompting the 6LoWPAN MAC layer to prepare for a BB signal. At this time, the 6LoWPAN MAC layer can be in any state defined by the IEEE 802.15.4 standard such as backing off and sensing the channel, so the MAC layer needs to wait until the next beacon cycle begins and the beacon can be used to carry a BB signal. Once the 6LoWPAN MAC layer is prepared to transmit a beacon, a BB reply is sent back to the WLAN application layer, informing that the 6LoWPAN MAC layer is about to transmit a beacon. After that, the BB signal is injected into the beacon packet and disseminated to the 6LoWPAN end devices.

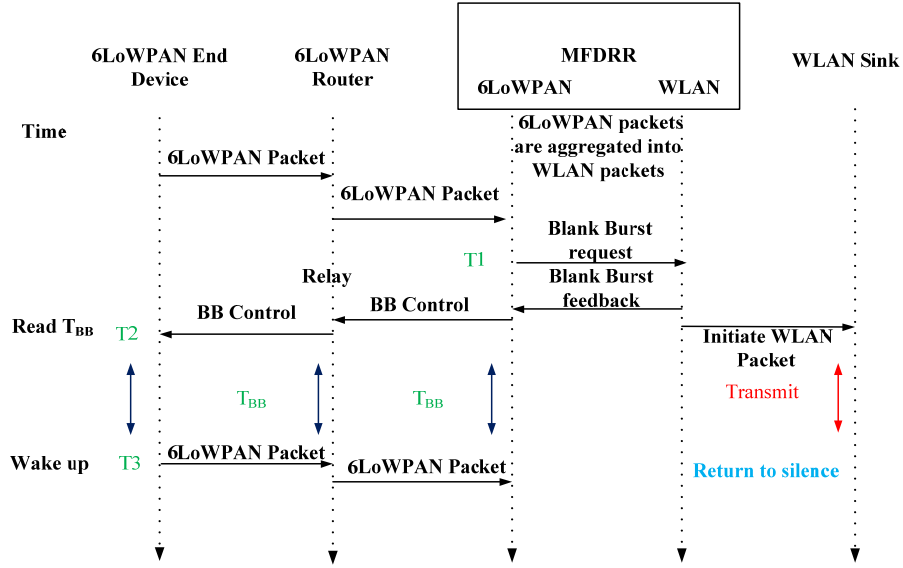

Fig. 8 The Blank Burst process

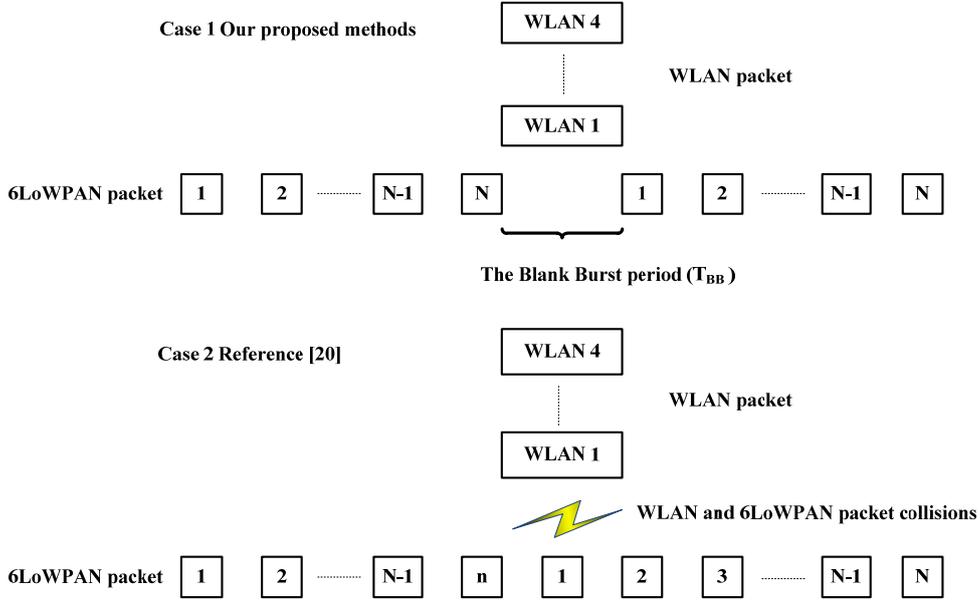

Fig. 9 Difference between our methods and work [20] in the literature

Meanwhile, the MFDRR has a timer that records the time it takes to transmit the BB notification from the MFDRR to the 6LoWPAN end devices. As the network is synchronised by the beacon emitted by the MFDRR and the Router, the duration for a beacon transmission time from the MFDRR to the end devices is fixed. Therefore, when the timer runs out, it means that the beacon has reached the end devices at time T2 via the router. Upon receiving the beacon at time T2, the end devices extract the BB signal, read $T_{BB}$, and remain silent during this period, which is calculated in equation (3). The WLAN Distributed Coordination Function (DCF) is used to transmit the WLAN packets. $T_{DIFS}$ denotes the DCF inter-frame spacing; $T_{backoff\_min}$ denotes the minimum back-off delay; the parameter $L_{agg}$ is one aggregated packet transmission time, including the headers and payloads; and $T_{SIFS}$ denotes the short interframe spacing. $T_{ACK}$ represents the acknowledgement packet transmission delay. N denotes the number of WLAN packets transmitted per BB duration. It is noted the back-off duration is the minimum back-off time because there is no other WLAN device competing for the channel. In other words, after a WLAN packet is transmitted, the next packet can be sent without contending for the channel. Meanwhile, the stored 6LoWPAN packets are aggregated into N WLAN packets, which in turn are transmitted via the WLAN sink during the $T_{BB}$ period. After the completion of this Blank Burst at time T3, the 6LoWPAN end devices wake up and resume sending data packets, and then a new BB cycle begins. Fig. 9 shows the difference between our proposed Blank Burst method and the existing work in the [20], which is used to compared to our method. It can be seen that case 1 reserves the blank burst period for WLAN packets, whereas case 2 does not. As a result, the existing work [20] still encounters the inter-network collisions; namely, 6LoWPAN and WLAN packet collisions.

$$T_{BB} = N \times (T_{DIFS} + T_{backoff\_min} + L_{agg} + T_{SIFS} + T_{ACK}) \tag{3}$$

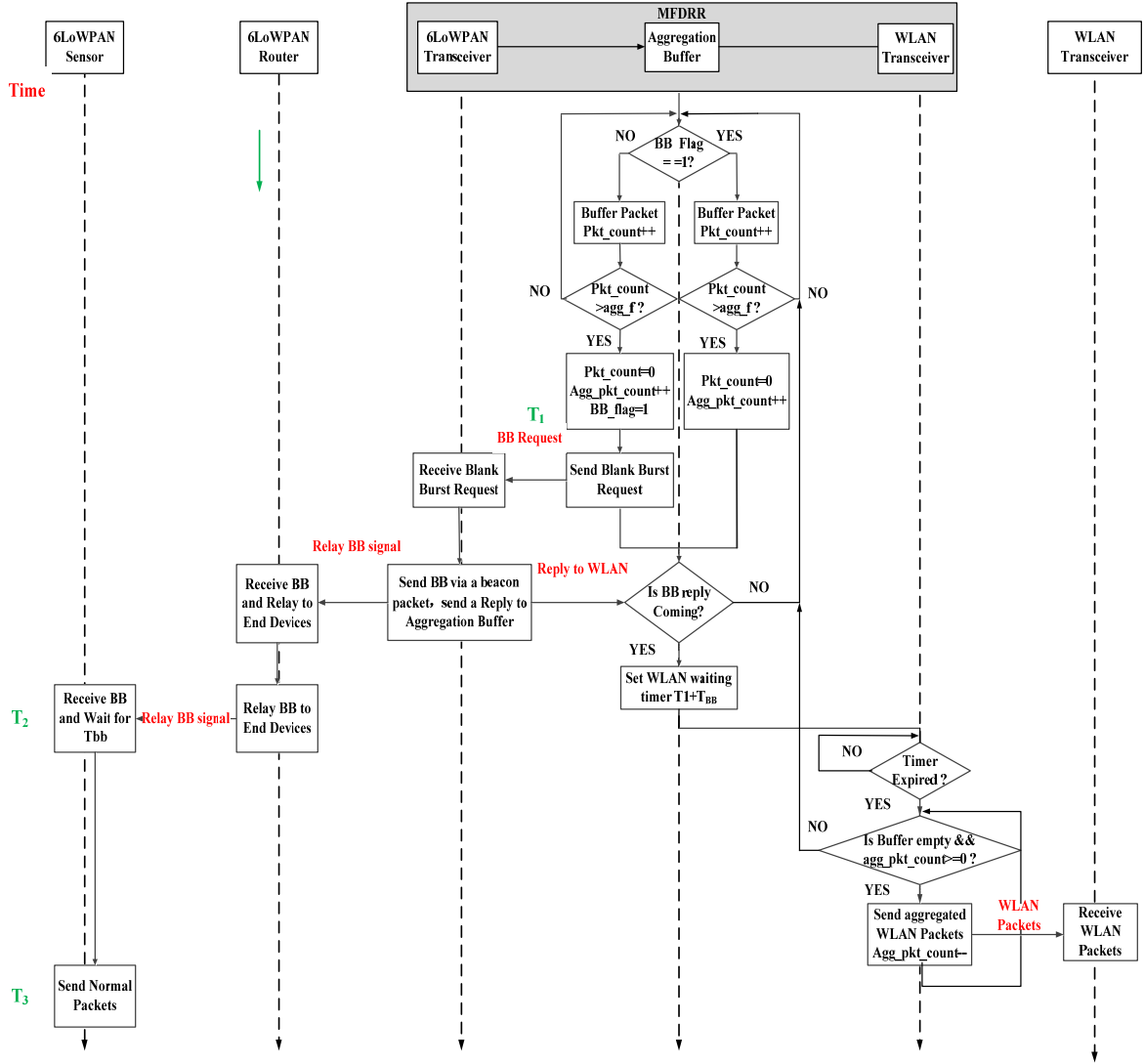

Fig. 10 Proposed aggregation factor-based Blank Burst (BB) algorithm

As can be seen in Fig. 10, the Blank Burst process is illustrated by the algorithm. The collaboration between each entity is illustrated using the time-based flow chart. The algorithm is executed chronologically as signalled by the green arrow and this explains the specific details of the tasks required by each type of the device. The algorithm is also driven by an aggregation factor, which is a pre-defined threshold, allowing the 6LoWPAN packets to enter the buffer before triggering the BB algorithm. Once the 6LoWPAN payloads continue to enter the aggregation buffer and the number of the 6LoWPAN payloads reaches the aggregation factor agg_f, the BB algorithm is triggered. After that, a Blank Burst request is sent to the 6LoWPAN MAC layer of the MFDRR, which can be in any state, and thus the network needs to wait until the 6LoWPAN MAC layer is ready to send the next beacon. When the beacon is ready, the value of the BB period is wrapped in a beacon frame and transmitted to the end devices. Meanwhile, a reply is sent to the aggregation buffer to trigger the timer set for WLAN packet transmissions. When the timer goes off, the WLAN packet transmissions begin, while the 6LoWPAN end devices receive the signalling beacon and remain silent for the BB period to avoid the inter-network collisions.

## 5 Lifetime-based BB algorithm design

### 5.1 QoS-Aware Lifetime-based Queuing Design

In this section, the lifetime-based BB algorithm is proposed to improve the performance of the previous aggregation factor-based BB algorithm and mitigate the inter-network collisions. To meet stringent delay requirements of different M2M applications, it is necessary to assign the packets with different delay budgets (lifetime) in the network. If the lifetime is exceeded before the packet reaches the data sink, it is no longer considered useful and should be destroyed. Different types of packets should be put into different buffers or queues waiting to be transmitted. As presented in Fig. 11, in this paper, we consider three queues located at the application layer of the cluster head (the MFDRR) to differentiate the three types of traffic corresponding to three types of applications. The number of queues can vary depending on the numbers of applications. Each queue maintains a timer recording the current minimum lifetime. Specifically, as one packet enters the queue, the lifetime of the packet is retrieved and compared with the other lifetimes in the queue to find the minimum value. Since three traffic queues all have its own minimum lifetime values, it must determine the smallest value to trigger the BB algorithm.

More precisely, as a new packet enters the queue, the scheduler takes two steps before triggering the BB algorithm. The first is to compare the current minimum lifetime value in the one queue; the second is that a scheduler compares three minimum lifetime values computed from the three traffic queues to determine the smallest lifetime, which will trigger the BB algorithm. In addition, since a large number of 6LoWPAN packets are transmitted, it is necessary to aggregate them into IEEE 802.11 packets to improve channel efficiency. All of the packets in the three queues are packed into IEEE 802.11g payloads and transmitted by the BB algorithm. Further, given that the maximum packet WLAN MTU length is 2304 bytes, the aggregation factor is used to control the number of the 6LoWPAN packets in one IEEE 802.11 payload. For example, a large aggregation factor allows more 6LoWPAN packets to be packed than a small aggregation factor. A large aggregation factor generates fewer IEEE 802.11 packets, so the channel utilisation is higher. Fig. 11 shows three cases of the lifetime values. The blue, red and green blocks represent the different lifetime values of an incoming packet $T_2$, the current absolute time is $T_0$ and the current shortest lifetime maintained by the queue is $T_1$. Another safety margin $T_{margin}$ is also set to avoid timeout when the BB signal is being processed.

In case 1, the new packet lifetime is $T_2 < T_0$, which means that the lifetime has expired so that this packet must be dropped. Also, if $T_0 < T_2 < T_{margin}$, the lifetime is later than the current time, but smaller than the minimum time reserved for the Blank Burst triggering $T_{margin}$. In other words, since the BB signalling needs a margin area, falling behind this region means that this packet will expire before the Blank Burst algorithm is triggered, so this packet also should be dropped.

In case 2, if $T_0 < T_{margin} < T_2 < T_1$, it means that the new lifetime is shorter than the current lifetime, and thus the new lifetime should replace the current lifetime and become the new minimum lifetime. The previous timer should be cancelled. This case updates the timing to trigger the BB algorithm.

In case 3, if $T_1 < T_2$, it means that current lifetime is the minimum lifetime, there is no need to update the current lifetime for this queue. Therefore, this packet should be stored in the queue.

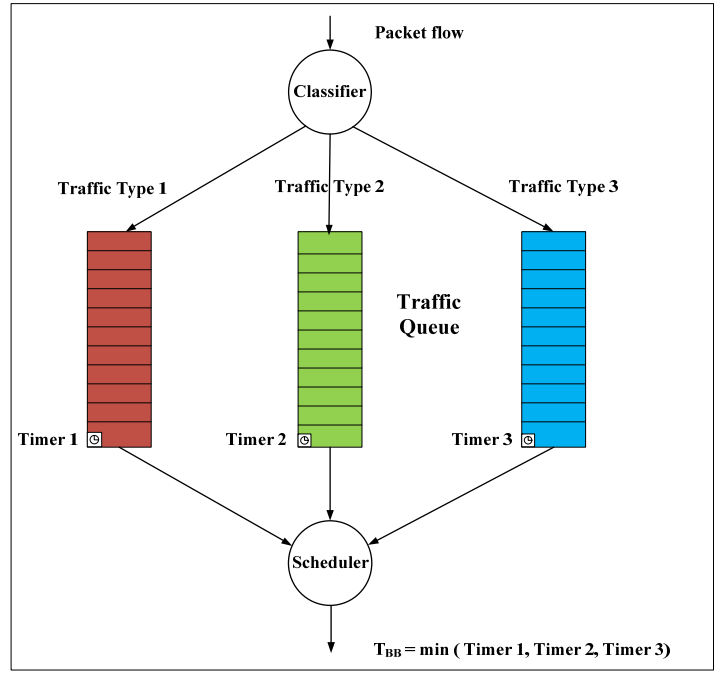

Fig. 11 MFDRR application layer queue structure

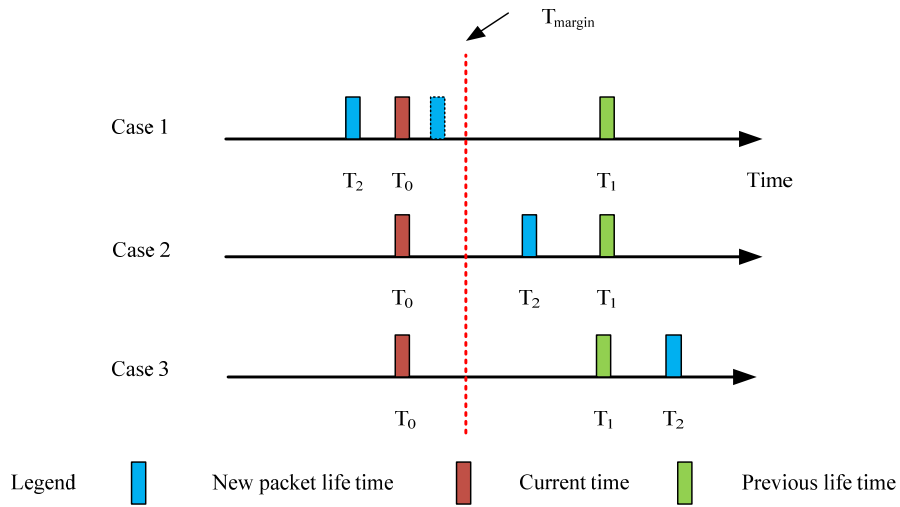

Fig. 12 Blank Burst algorithm is triggered based on lifetime

These three cases are only used for one type of traffic, and each packet in the queue must maintain the current minimum lifetime to make sure that an urgent request can be handled immediately. Before triggering the BB algorithm, it needs to determine the shortest lifetime among the three minimum lifetime values as the real minimum lifetime. To do this, the three minimum values in each queue are compared using a customised function and the shortest one is used as the BB algorithm timer. In addition, the aggregation factors are used to regulate the number of packets wrapped in IEEE 802.11 packet payloads.

Figure 12 shows the timing instants at which that the Blank Burst algorithm is launched. The minimum lifetime in the network can elapse in any point in time within the network operating hours, meaning that the Blank Burst algorithm can be triggered at any time as well, so the number of 6LoWPAN packets could not be an integer multiple of the aggregation factor, so at least two cases must be considered.

$$N_1 < \text{agg\_factor} \tag{4}$$

$$N_1 \geq \text{agg\_factor} \tag{5}$$

In equation (4), $N_1 <$ agg_factor, where $N_1$ represents the number of 6LoWPAN packets packed into one IEEE 802.11 packet and the number of IEEE 802.11 packets $N_2=1$. In equation (5), $N_1 \geq$ agg_factor, the 6LoWPAN packets are packed into several IEEE 802.11g packets. This case can be divided into two sub-cases: (a) $N_1$ mod agg_factor=0 and (b) $N_1$ mod agg_factor$\neq$0. The first case indicates that N1 is an integral multiple of agg_factor and the number of the aggregated IEEE 802.11 payloads is $N_2=N_1$/agg_factor. In equation (5), the number of the aggregated IEEE 802.11g payloads is $N_2=[N_1$/agg_factor$]+1$, as the remaining packets will be packed into a separate IEEE 802.11 packet and then transmitted. During the WLAN packet transmissions, the combination of the packet lifetime and aggregation factor ensures a timely transmission and channel efficiency. Specifically, a packet lifetime guarantees the delay budget and triggers the BB algorithm, while the aggregation factor regulates the number of the IEEE 802.11 packets. Therefore, the lifetime-based BB algorithm is more complicated than the previous one, which only uses the aggregation factor to trigger the BB algorithm. Selecting the minimum lifetime from the three queues ensures that the urgent packet has the chance to be transmitted to the data sink to fulfil its QoS requirements. In addition, after the BB algorithm is triggered, a packet with a small lifetime must be transmitted together with the remaining packets in the queue. The urgent queue needs to be emptied first. If the Blank Burst duration is long enough, then the other two queues can also be emptied depending on the number of packets in those queues and the BB transmission duration.

## 5.2 Enhanced Blank Burst Algorithm Design

This section presents an enhanced algorithm that employs the packet lifetime to trigger the BB algorithm. The new algorithm is based on the previous one; that is, the BB signal is emitted from the MFDRR MAC layer, then is relayed to 6LoWPAN devices and finally suspends the 6LoWPAN transmissions, thus leaving this period for the MFDRR 802.11 interface. With the algorithm, the 6LoWPAN and IEEE 802.11 transmitters can send packets without interfering with each other. Fig. 13 presents the three stages of the lifetime-based BB algorithm. The first stage is that as a packet comes into the application layer of the MFDRR, the packet is stored in the queue that stores the same type of packets by a classifier. Meanwhile, the lifetime of this packet is retrieved and compared using algorithms 1 and 2. The second stage is the BB signal generation, in which the signal is embedded in one field of the beacon packet, then transmitted and relayed to the 6LoWPAN end devices. The third stage is when the signal is received by the 6LoWPAN devices, it stops the transmissions and allows the IEEE 802.11g transmissions with algorithm 3 in the BB period. More precisely, the number of the aggregated WLAN payloads is calculated using the aggregation factor and the number of the 6LoWPAN packets. When the three queues are flushed, the queue with the minimum lifetime must be flushed first, and subsequently the second and third non-urgent queues to ensure the QoS of M2M applications. To better interpret the algorithm, three algorithms 1, 2 and 3 are explained in pseudo-code as they are repeatedly invoked within the improved BB algorithm.

As shown in Fig. 13, when a new packet enters the queue, the packet type is determined by the classifier. Algorithm 1 is used to identify the shortest lifetime of this queue by measuring the lifetimes of all the queued packets; algorithm 2 uses the lifetime value to trigger the BB algorithm based on the results obtained from algorithm 1, as shown in the blue rectangle. Once the minimum lifetime is obtained, the BB timer is set. This means the QoS of the network relies on this timer, and when it expires, the packets are transmitted using the enhanced BB algorithm, which means the stringent QoS requirements of most applications are met. The WLAN standard supports much higher transmission rates than the 6LoWPAN standard, so it is necessary that the WLAN interface of the MFDRR should flush the three queues as much as possible. In addition, after the expiration of the timer, the BB signal is emitted from the MFDRR MAC layer and forwarded to the 6LoWPAN devices. Upon receiving the signal, the 6LoWPAN devices wait for a BB period of time for the WLAN transmissions. As for the IEEE 802.11 side of the MFDRR, the IEEE 802.11 interface attempts to flush the three queues according to the packet priorities. The queue with the highest priority is flushed first. Combined with the aggregation factors, it can be seen that Algorithm 3 in the green rectangle is repeatedly revoked to transmit different numbers of WLAN packets. After the end of the WLAN transmissions, the cycle begins.

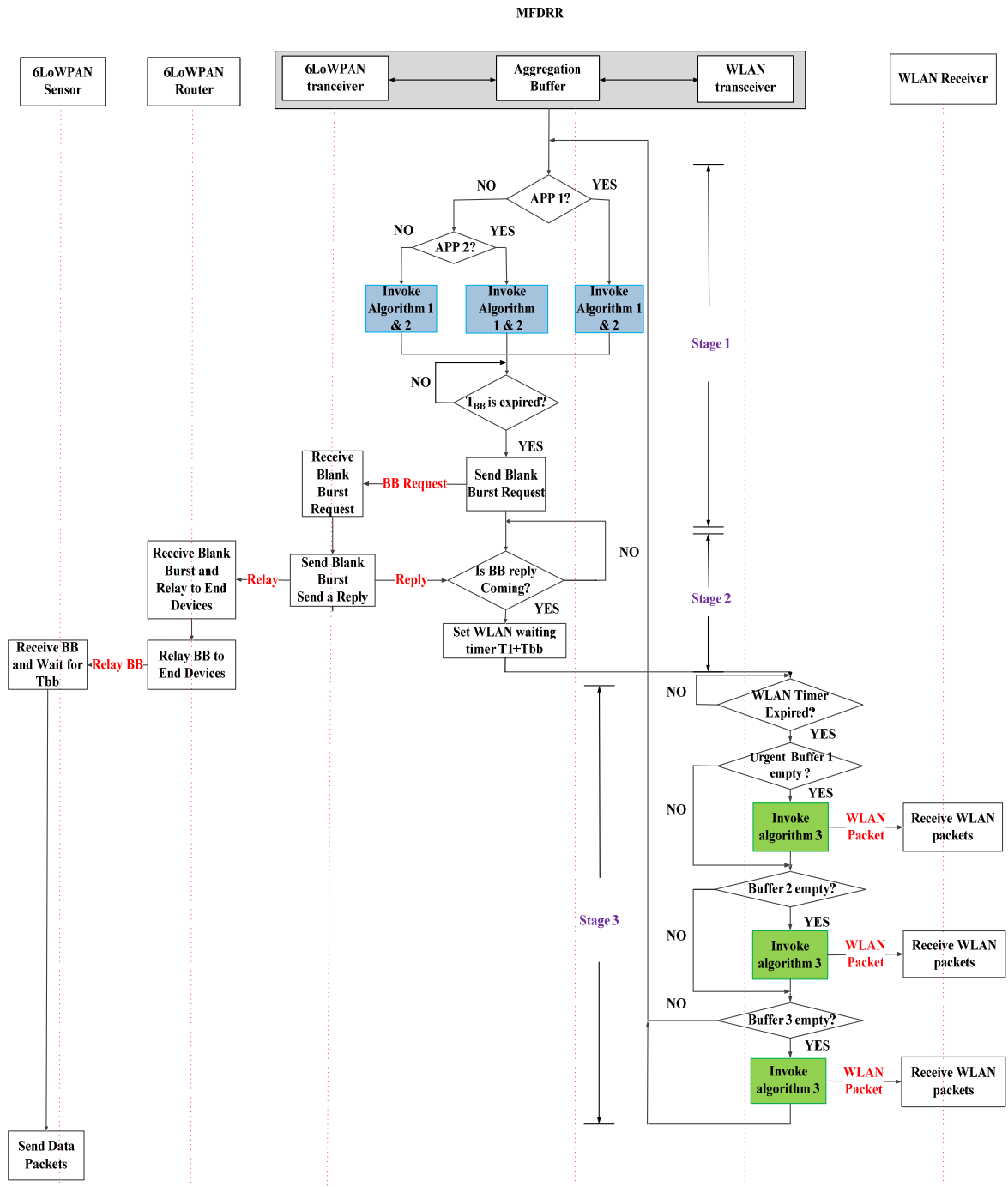

Fig. 13 Flow chart of the improved Blank Burst algorithm

## Algorithm 1

|   | Algorithm 1: Scheduler to identify the minimum lifetime value |
|---|---|
|   | **Input**: deadline_1, deadline_2, deadline_3 |
| 1 | **Output**:BB_deadline |
| 2 | **Compare throught** deadline_1 deadline_2 and deadline_3 **Then** |
| 3 | **Return** BB_deadline:=**min** (deadline_1, deadline_2, deadline_3); |
| 4 | **End** |

## Algorithm 2

**Algorithm 2: Lifetime Scheduling**

| | |
|---|---|
| 1 | **Input**: $T_2$, $T_0$, $T_{margin}$ |
| 2 | Invoked by each packet queue to determine the newest minimum lifetime |
| 3 | **If** Mutex==0 **Then**     // no Blank Burst signal has been triggered. |
| 4 |    **If** deadline_1 ==0 **Then** //First time no values given to deadline_1 |
| 5 |         **If** $T_2 - T_0 < T_{margin}$ **Then** |
| 6 |            Drop this packet; |
| 7 |         **End** |
| 8 |         **If** $T_2 - T_0 == T_{margin}$ **Then** |
| 9 |            Trigger Blank Burst (BB) algorithm and insert the packet into the queue; |
| 10 |         **End** |
| 11 |         **If** $T_2 - T_0 > T_{margin}$ |
| 12 |            Deadline_1:= $T_2$; |
| 13 |            BB_deadline:= invoke **Algorithm 1** (deadline_1, deadline_2, deadline_3); |
| 14 |            **Set** BB timer $T_{timer}$=BB_deadline;// Obtain the new minimum lifetime |
| 15 |            Insert the packet in the queue; |
| 16 |         **End** |
| 17 |    **Else If** (deadline_1!=0) **Then** |
| 18 |         **If** deadline_1 < $T_2$ **Then** |
| 19 |            Insert the packet in the queue; |
| 20 |         **End** |
| 21 |         **If** deadline_1 > $T_2$ && $T_2 - T_0 == T_{margin}$ |
| 22 |            Cancel the previous timer $T_{timer}$ |
| 23 |            Trigger BB algorithm and insert the packet in the queue; |
| 24 |         **End** |
| 25 |         **If** deadline_1 > $T_2$ && $T_2 - T_0 > T_{margin}$ |
| 26 |            Deadline_1:= $T_2$; |
| 27 |            BB_deadline:= invoke **Algorithm 1** (deadline_1, deadline_2, deadline_3); |
| 28 |            Cancel the pervious timer $T_{timer}$ |
| 29 |            Insert the packet in the queue; |
| 30 |            Set BB timer $T_{timer}$:=BB_deadline; |
| 31 |         **End** |
| 32 |         **If** deadline_1 > $T_2$ && $T_2 - T_0 < T_{margin}$ |
| 33 |            Drop this packet; |
| 34 |         **End** |
| 35 |    **End** |
| 36 | **Else If** Mutex!=0;  // a Blank Burst signal has been triggered |
| 37 |    Insert the packet into the queue |
| 38 | **End** |

## Algorithm 3

**Algorithm 3: Count the number of the aggregated packets and send them out**

**Input**: pkt_count, agg_factor, agg_pkt_count

| | |
|---|---|
| 1 | **If** pkt_count ≤ agg_factor **Then** |
| 2 |    Remove packets from the aggregation buffer; |
| 3 |    Create an aggregated IEEE802.11g payload; |
| 4 |    Send the aggregated payload out; |
| 5 | **Else if** pkt_count ≥ agg_factor **Then** |
| 6 |       **If** pkt_count mod agg_factor == 0 **Then** |
| 7 |         agg_pkt_count:= pkt_count/agg_factor ; |
| 8 |         While agg_pkt_count !=0 do |
| 9 |         Obtain agg_factor packets from the buffer; |
| 10 |         Create an aggregated WLAN payload; |
| 11 |         Send this aggregated payload out; |
| 12 |       **End** |
| 13 |     **Else** |
| 14 |         agg_pkt_count:= $\lfloor$ pkt_count/agg_factor $\rfloor$ +1 |
| 15 |         While agg_pkt_count !=0 do |
| 16 |         Obtain agg_factor packets from the buffer; |
| 17 |         Create an aggregated WLAN payload; |
| 18 |         Send this aggregated payload out; |
| 19 |     **End** |
| 20 | **End** |

# 6 Performance analysis of the proposed two interference mitigation techniques

## 6.1 Simulation Setup

The performances of the aggregation-factor based and lifetime-based BB algorithms are evaluated in this section, and the key simulation parameters are listed in Table 1. We conducted the simulation multiple seed values and the plotted results plotted with a 95% confidence interval. The lifetime requirements for sensors and meter reading represented in the table are sourced from [24]. The simulation consists of two parts. In the first part, the algorithm proposed in a similar research work [20] was used to compare against our proposed algorithms. In the second part, the meter reading and sensor traffic are included to show that the proposed algorithms can meet basic QoS requirements for these two M2M applications.

Table 1 The key parameters for the lifetime-based BB algorithm

| Group Name | Parameter | | Value |
|---|---|---|---|
| **Network** | Hop count | | 3 |
| | Number of nodes | | 293 |
| | Standard | | 6LoWPAN and IEEE 802.11g |
| | Operating Frequency | | 2.4 GHz |
| | 6LoWPAN channel | | 12 |
| | IEEE 802.11g channels | | 1 and 6 |
| **Propagation model** | Free space path loss | | |
| **MFDRR** | 6LoWPAN | BO | 4 |
| | | SO | 3 |
| | | Transmit Power | 1.8 mW |
| | WLAN | Transmit Power | 100 mW |
| | | Packet Size | 1200 bytes |
| | | Aggregation Factor | 25 |
| | | Blank Burst algorithm safety margin (s) | 0.2 s |
| **Router** | BO | | 4 |
| | SO | | 2 |
| | Transmit Power | | 1.8 mW |
| **End device** | Packet size | | 64 bytes |
| | Packet generation | | Exponentially distributed |
| | Transmit Power | | 1 mW |
| | Packet inter-arrival rate | | 0.5, 1, 1.3, 1.5, 1.7, 2 pkts/s |
| **Target Applications** | Sensor end-to-end delay | | 0.6 s |
| **QoS Requirements** | Meter Reading end-to-end delay | | 900 s |

## 6.2 Performance Analysis of the Proposed Algorithms

In this section, the effectiveness of several interference mitigation algorithms are tested. In the four-MFDRR area network, four scenarios are considered. This first scenario does not adopt the proposed algorithms, so all the 6LoWPAN devices are subject to inter-network collisions. The second scenario uses the algorithm proposed in [20], which is an adaptive packet aggregation algorithm (referred to as agg in the figure legend). It aggregates the 6LoWPAN packets into a WLAN packet to reduce the number of the transmitted packets, thus alleviating the impact of inter-network collisions. The difference between this method in the literature and our proposed algorithms is that the existing method does not use the Blank Burst period to protect the 6LoWPAN packets, so this method still experiences the inter-network collision. The third scenario uses the aggregation factor-based algorithm proposed in this paper, and the fourth scenario employs the lifetime-based BB algorithm also proposed in this paper.

Figure 14 shows the packet delivery rates for the four scenarios just discussed above. It can be seen that the scheduling algorithms have significantly improved the performance of the area network compared with the scenario one where no scheduling algorithm was used. Specifically, It can be seen that the packet delivery rate (PDR) in the first scenario declined from 28% to 8% with the increasing incoming packet rate. In contrast, the PDR decreased from 97% from to 68% when the BB algorithms were used. In particular, the agg algorithm proposed in [20] has a 20% lower PDR than those of the aggregation factor and lifetime-based BB algorithms at high incoming packet rates. It is because the adaptive aggregation algorithm aggregated 6LoWPAN packets with no guard time. As a result, as the load increased, the WLAN packets more severely affected the 6LoWPAN network, thus resulting in a higher number of packet losses. In contrast, the aggregation-factor and lifetime-based algorithms not only used the aggregation to reduce the number of packets, but used a short BB period to protect the 6LoWPAN network from the WLAN transmissions. It is noted that the lifetime-based algorithm had a slightly higher PDR than the aggregation factor BB algorithm at the high traffic load, which is higher than 1.3 pkts/s. The simulation results show the effectiveness of the proposed BB algorithms in mitigating the inter-network collisions.

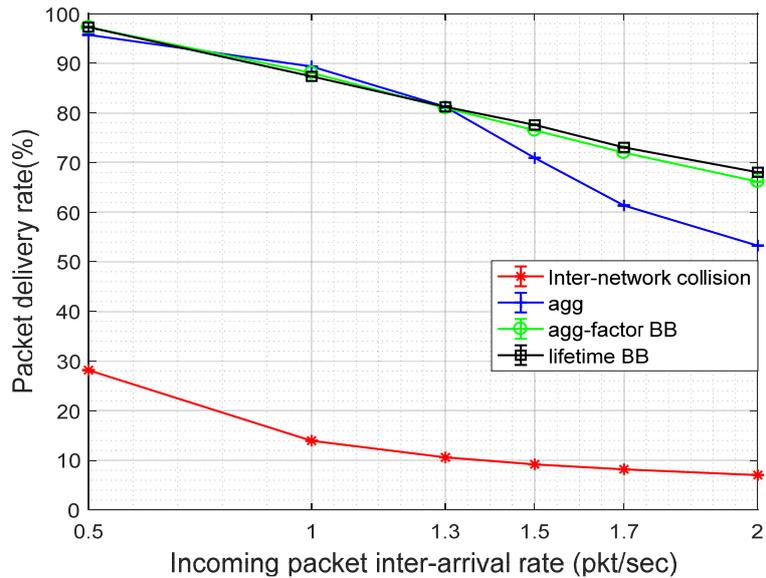

Fig. 14 The packet delivery rates of four MFDRRs with the proposed algorithms

Figure 15 shows the end-to-end delay with four MFDRRs and the proposed algorithms. To show more details for the adaptive aggregation algorithm, aggregation factor-based and lifetime-based BB algorithms, a magnified picture is presented in Fig. 16. It can be seen that the end-to-end delay of the case without any algorithms slowly rose to 7 s and then climbed to 36s, whereas the other three algorithms maintained relatively low delays, with only the adaptive aggregation algorithm rising to 7.8s at high traffic loads of 1.5 pkts/s. The reason was the inter-network collisions that caused the increased router's MAC queuing delay, which was the major delay component for the end-to-end delay. Fig. 16 presents the end-to-end delays for the adaptive aggregation algorithm, aggregation factor-based BB and lifetime-based BB algorithms. It can be observed that the delay of the adaptive aggregation algorithm increased as the load increased beyond 1.5 pkts/s. The rationale is that this algorithm does not have the guard time to protect the 6LoWPAN transmissions, and thus the WLAN interfaces cause longer router queuing delay. In contrast, the end-to-end delays of the aggregation factor-based and lifetime-based algorithms were maintained at 0.8s, which proves the effectiveness of the proposed algorithms.

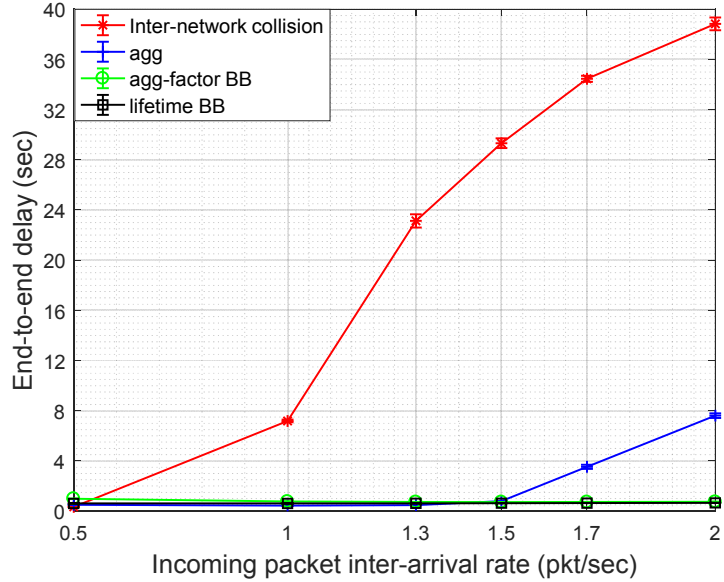

Fig. 15 End-to-end delay with four MFDRRs

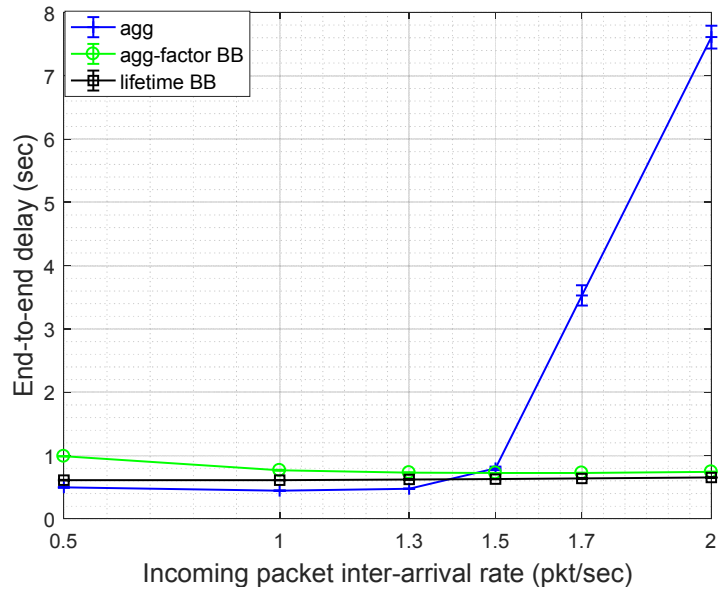

Fig. 16 Magnified end-to-end delay

Figure 17 and Figure 18 show the end device queuing delay and router queuing delay. It can be seen that the inter-network collisions negatively affected the 6LoWPAN end devices. This was because the staggered link design scheduled the WLAN transmissions within the superframe, in which the communication between the end devices and routers occurs. Despite using the BB algorithm, the end devices were still subject to the inter-network collisions, so the figure shows the slightly higher delays than those of the no algorithms involves scenario and the adaptive aggregation scenario. It is noted that the adaptive aggregation algorithm had no Blank Burst period to protect the 6LoWPAN devices, so the routers experienced the rising delays instead of the end devices. On the other hand, the proposed BB algorithms showed a clear advantage of the end-to-end delay and maintained the low delays with the increased traffic loads. Another reason for such low delays is that the BB period protected the 6LoWPAN transmissions, so the packets were not subject to the inter-network collisions. Therefore, the proposed two algorithms had lower queuing delays.

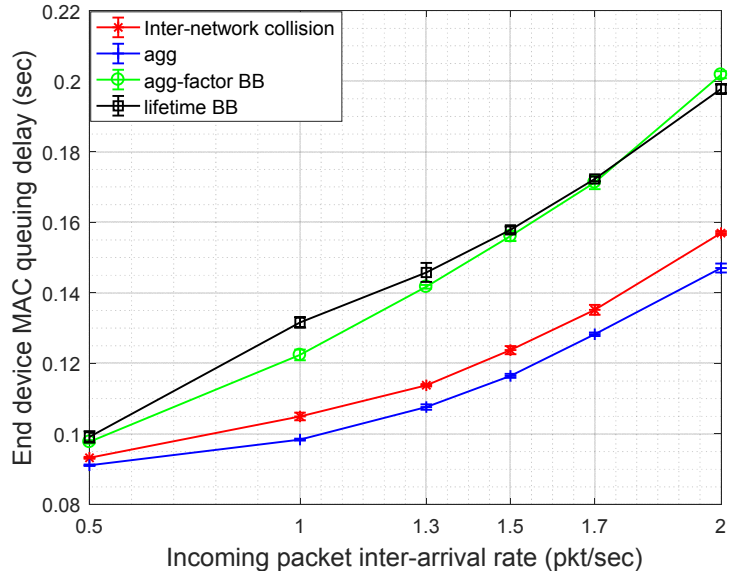

Fig. 17 End device MAC queuing delay

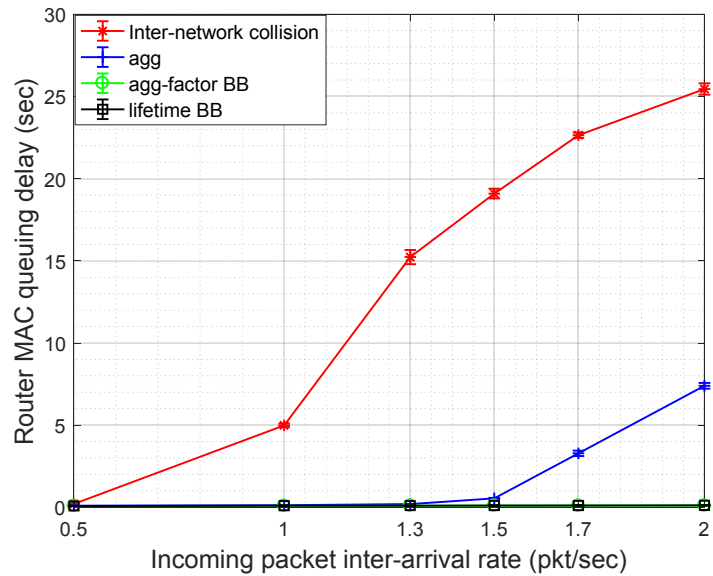

Fig. 18 Router MAC queuing delay

Figure 19 shows the total throughput of the four MFDRRs. It is obvious that the throughput of scenario one without the proposed algorithms stayed as low as 35 pkts/s due to the adverse impacts of the inter-network collisions. In contrast, in the other three cases the network witnessed a rising throughput up to 350 pkts/s. The reason was that the packet aggregation technique significantly reduced the number of the WLAN packets transmitted from the MFDRRs, thus minimizing the negative impacts of the inter-network collisions. In addition to the packet aggregation, the aggregation factor-based algorithm and lifetime-based algorithm used the BB techniques to avoid the direct collisions with the WLAN packets. However, the adaptive aggregation algorithm does not use this BB period, resulting in the degradation of the QoS due to the inter-network collisions. This can be reflected in Fig. 20 that shows the total packet losses, where the adaptive aggregation algorithm shows more packet losses compared to the aggregation factor-based BB and lifetime-based BB algorithms.

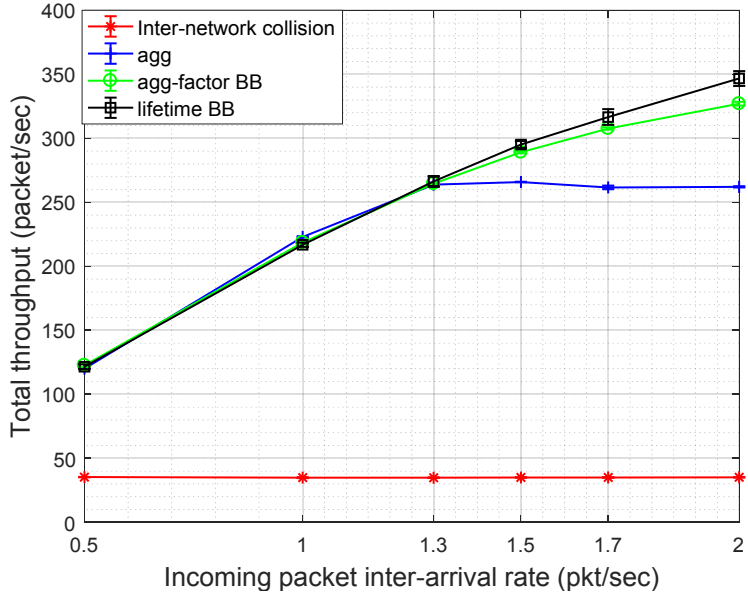

Fig. 19 Total MFDRRs' throughput

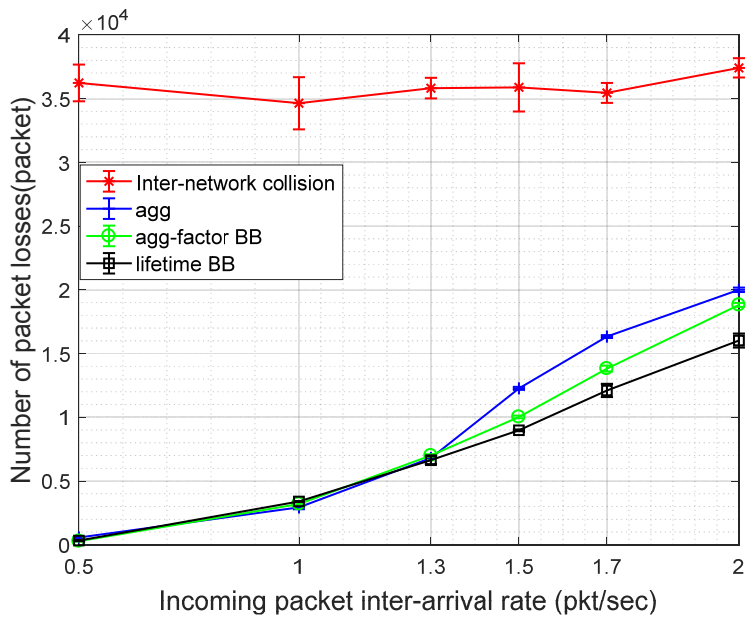

Fig. 20 Total packet losses

Moreover, we introduce two types of traffic to test the performances of the proposed lifetime-based BB algorithms. Specifically, we examine the sensor traffic and smart metering traffic with the QoS of requirements. The sensor traffic is bounded by the end-to-end requirement of 1s for M2M applications such as smoke detectors, while the meter reading traffic delay requirement is 900s. In the simulation, 32 out of the 64 6LoWPAN nodes in one area network are sensor nodes, while the other 32 nodes are meter reading nodes.

The packet delivery rates of the sensor traffic and meter reading traffic are presented in Fig. 21, in which the lifetime-based Blank Burst was used. The sensor packet delivery rate gradually dropped from 98% to 60% as the traffic load increased, while the meter reading packet delivery rate remained stable at nearly 99% over the simulation time. The difference is that the sensor traffic, due to its higher packet inter-arrival rate, had a higher chance of colliding with the WLAN packets. The results also reveal that the access networks such as 6LoWPAN and WLAN networks tend to experience more packet collisions when traffic loads increase. Therefore, the network must maintain lower traffic loads to meet the QoS requirements of M2M applications.

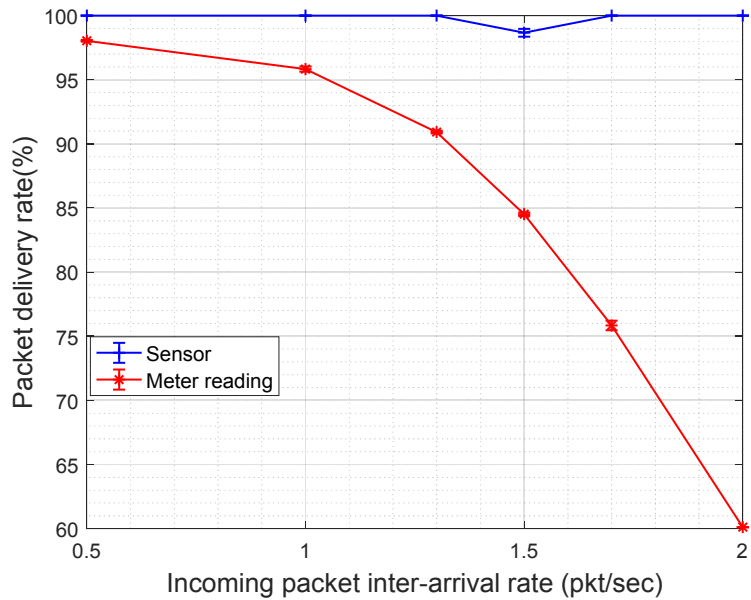

Fig. 21 Packet delivery rate with the sensor and meter reading traffic

The end-to-end delays for the two types of traffic are illustrated in Fig. 22. Both delays were quite stable around 0.56s to 0.6s even at traffic loads of 2 pkts/sec. The meter reading traffic load was not high, so the end-to-end delay levelled at 0.6s. Although the end-to-end delay of the sensor traffic was not high, it experienced packet losses at the high traffic loads due to the intra-network collisions. These results prove that the proposed heterogeneous area network meets the end-to-end delay requirements of the basic M2M applications.

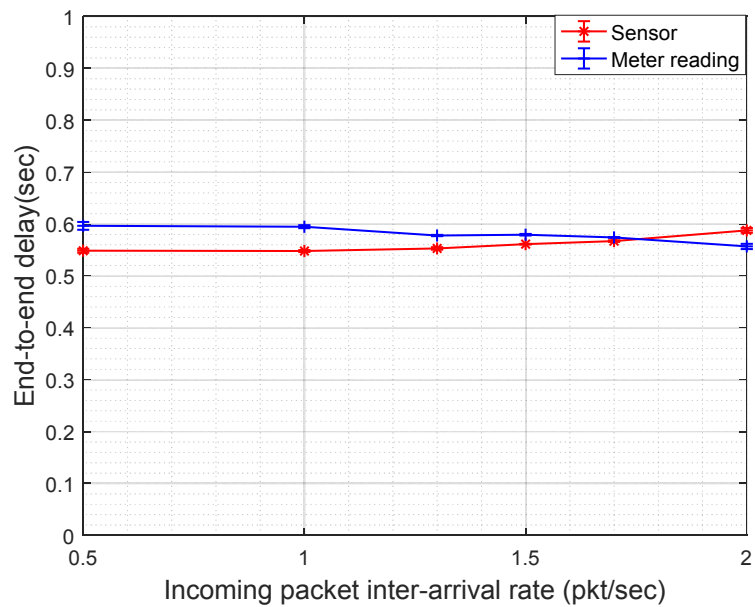

Fig. 22  End-to-end delay with the sensor and meter reading traffic in one area network

The number of packet losses for the sensor and meter reading traffic for the lifetime-based algorithm is depicted in Fig. 23. It can be seen that the sensor packet losses increased from zero to ten packets as the traffic load increased, whereas the meter reading loss is always zero. This is because the increased sensor loads slightly increased the queuing delay of the sensor traffic, so a small number of packets, which have exceeded their lifetime, were dropped by the MFDRR. In contrast, the meter reading traffic losses were low because the meter reading loads were low compared to the sensor traffic. The result shows that these two types of traffic can be well supported by the proposed area network.

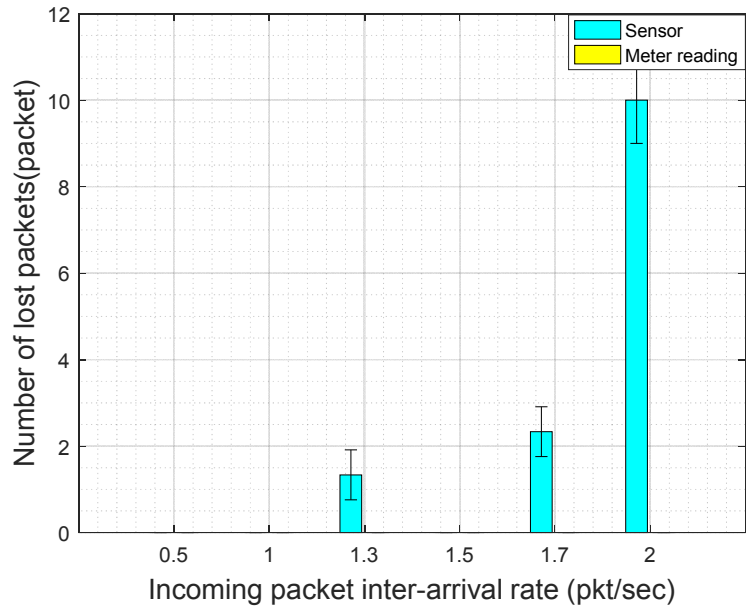

Fig. 23 Mean sensor and meter reading packet losses in one area network

### 6.3 Discussion

The results show superiority to the existing method proposed in [20] and would help to gain advantages in 5G communications environments. This is because various types of networks are emerging as a heterogeneous network in the 5G era, and network convergence plays a critical role in assisting the smooth data transmission [25]. Many more of these data exchange could occur in the residential or industrial environment using the smaller cell technology [26], so the 2.4 GHz license-free band can be inundated with increasingly more types of devices and multi-layer networks. For example, when the Massive Output Massive Input (MIMO) system [27] is used in the building blocks in which smart meters with sensors collecting data from home smart appliances using the 2.4 GHz band, the proposed algorithms would be useful in terms of mitigating the inter-network collisions and enhancing the network performance.

## 7 Conclusion

In this paper, we presented two interference mitigation techniques for a large-scale dense heterogeneous wireless area network. The proposed algorithms show significant gains in terms of the packet delivery rate, end-to-end delay and throughput compared to the existing algorithm in the literature. In particular, the lifetime-based BB algorithm greatly reduces the number of inter-network collisions and prevents the network from being interfered by the inter-network collisions from its own MFDRR and its neighbouring MFDRRs. The simulation results show that both the algorithms can mitigate the inter-network collisions in a large dense heterogeneous area network while maintaining the basic QoS for M2M applications. They also show superiority to the existing method proposed in the literature, which gains an advantage in 5G M2M environments in which short-rang heterogeneous area networks would play an important role.

[1]     L. Atzori, A. Iera, and G. Morabito, "The internet of things: A survey," *Computer networks,* vol. 54, pp. 2787-2805, 2010.

[2] F. Chen, X. Tong, E. Ngai, and F. Dressler, "Mode switch—Adaptive use of delay-sensitive or energy-aware communication in IEEE 802.15. 4-based networks," in *Mobile Adhoc and Sensor Systems (MASS), 2010 IEEE 7th International Conference on*, 2010, pp. 302-311.

[3] M. Talebi, C. Papatsimpa, and J.-P. M. Linnartz, "Dynamic performance analysis of IEEE 802.15. 4 networks under intermittent Wi-Fi interference," in *2018 IEEE 29th Annual International Symposium on Personal, Indoor and Mobile Radio Communications (PIMRC)*, 2018, pp. 1-7.

[4] N. de Araújo Moreira, V. Toldov, R. Igual-Perez, R. Vyas, N. Miton, and L. Clavier, *Heterogeneous Networks: experimental study of interference between IEEE 802.11 and IEEE 802.15.4 technologies*, 2017.

[5] R. Casagrande, R. Moraes, C. Montez, A. Morales, and L. Rech, "Interference of IEEE 802.11 n Networks upon IEEE 802.15. 4-Based WBANs: An Experimental Study," in *2018 IEEE 16th International Conference on Industrial Informatics (INDIN)*, 2018, pp. 388-393.

[6] S. S. Wagh, A. More, and P. R. Kharote, "Performance evaluation of IEEE 802.15. 4 protocol under coexistence of WiFi 802.11 b," *Procedia Computer Science,* vol. 57, pp. 745-751, 2015.

[7] J. Han, J. Bang, and Y. Lee, "Transmission performance of wireless sensor networks in the presence of co-channel interference," in *2014 IEEE Wireless Communications and Networking Conference (WCNC)*, 2014, pp. 1956-1961.

[8] K. Vikram and S. K. Sahoo, "A Collaborative Framework for Avoiding Interference Between Zigbee and WiFi for Effective Smart Metering Applications," *ELECTRONICS,* vol. 22, pp. 48-56, 2018.

[9] K. Hong, S. Lee, and K. Lee, "Performance improvement in ZigBee-based home networks with coexisting WLANs," *Pervasive and Mobile Computing,* vol. 19, pp. 156-166, 2015/05/01/ 2015.

[10] S. Nishikori, K. Kinoshita, Y. Tanigawa, H. Tode, and T. Watanabe, "A cooperative channel control method of ZigBee and WiFi for IoT services," in *2017 14th IEEE Annual Consumer Communications & Networking Conference (CCNC)*, 2017, pp. 1-6.

[11] H. R. Chi, K. F. Tsang, K. T. Chui, H. S. Chung, B. W. K. Ling, and L. L. Lai, "Interference-Mitigated ZigBee-Based Advanced Metering Infrastructure," *IEEE Transactions on Industrial Informatics,* vol. 12, pp. 672-684, 2016.

[12] F. Inoue, et al. "Hybrid Station Aided Coexistence Scheme between Wireless PANs and Wireless LAN." IEICE Transactions on Fundamentals of Electronics, Communications and Computer Sciences 98.2 (2015): 578-588.

[13] A. Koubaa, M. Alves, and E. Tovar, "A two-tiered architecture for real-time communications in large-scale wireless sensor networks: research challenges," in *WIP Proceeding of the 17th Euromicro Conference on Real-Time Systems (ECRTS'05), Palma de Mallorca, Spain*, 2005.

[14] J. Leal, A. Cunha, M. Alves, and A. Koubaa, "On a IEEE 802.15. 4/ZigBee to IEEE 802.11 gateway for the ART-WiSe architecture," in *Emerging Technologies and Factory Automation, 2007. ETFA. IEEE Conference on*, 2007, pp. 1388-1391.

[15] Q. Li, D. Han, O. Gnawali, P. Sommer, and B. Kusy, "Twonet: large-scale wireless sensor network testbed with dual-radio nodes," in *Proceedings of the 11th ACM Conference on Embedded Networked Sensor Systems*, 2013, p. 89.

[16] M. Ha, S. H. Kim, H. Kim, K. Kwon, N. Giang, and D. Kim, "SNAIL gateway: Dual-mode wireless access points for WiFi and IP-based wireless sensor networks in the internet of things," in *Consumer Communications and Networking Conference (CCNC), 2012 IEEE*, 2012, pp. 169-173.

[17] H. Y. Tung, K. F. Tsang, H. C. Tung, V. Rakocevic, K. T. Chui, and Y. W. Leung, "A WiFi-ZigBee building area network design of high traffics AMI for smart grid," *Smart Grid and Renewable Energy,* vol. 3, p. 324, 2012.

[18] J. Wang and V. Leung, "Comparisons of home area network connection alternatives for multifamily dwelling units," in *New Technologies, Mobility and Security (NTMS), 2011 4th IFIP International Conference on*, 2011, pp. 1-5.

[19] P. L. Shrestha, M. Hempel, Y. Qian, H. Sharif, J. Punwani, and M. Stewart, "Performance modeling of a multi-tier multi-hop hybrid sensor network protocol," in *Wireless Communications and Networking Conference (WCNC), 2013 IEEE*, 2013, pp. 2345-2350.

[20] Y. Jeong, J. Kim, and S.-J. Han, "Interference mitigation in wireless sensor networks using dual heterogeneous radios," *Wireless Networks,* vol. 17, p. 1699, 2011.
[21] "[Online] Available: http://www.tinyos.net.."
[22] "open-zb--OpenSource Toolset for IEEE 802.15.4 and Zigbee,http://www.open.-zb.net."
[23] "OPNETWORK 2010 1508 Undertanding TCP/IP Model Internals and Interfaces https://enterprise14.opnet.com/4dcgi/CL_SessionDetail?ViewCL_SessionID=4211."
[24] M. Kuzlu, M. Pipattanasomporn, and S. Rahman, "Communication network requirements for major smart grid applications in HAN, NAN and WAN," *Computer Networks,* vol. 67, pp. 74-88, 2014.
[25] T. Han, X. Ge, L. Wang, K. S. Kwak, Y. Han, and X. Liu, "5G Converged Cell-Less Communications in Smart Cities," *IEEE Communications Magazine,* vol. 55, pp. 44-50, 2017.
[26] A. Cimmino, T. Pecorella, R. Fantacci, F. Granelli, T. F. Rahman, C. Sacchi*, et al.*, "The role of small cell technology in future smart city applications," *Transactions on Emerging Telecommunications Technologies,* vol. 25, pp. 11-20, 2014.
[27] H. Chu, L. Zheng, and X. Wang, "Super-Resolution mmWave Channel Estimation for Generalized Spatial Modulation Systems," *IEEE Journal of Selected Topics in Signal Processing,* 2019.